# Green Footprint by Cognitive Management of Opportunistic Networks

Marios LOGOTHETIS, George ATHANASIOU, Kostas TSAGKARIS, Panagiotis DEMESTICHAS
Department of Digital Systems, University of Piraeus,
80, Karaoli and Dimitriou Street, Piraeus, 18534, Greece
Email: {mlogothe, athanas, ktsagk, pdemest} @unipi.gr

*Abstract*— **The existing characteristics of the wireless networks nowadays, urgently impose the exploitation of flexible networking solutions that will offer increased efficiency in resource utilization and application Quality of Service (QoS) provisioning and at the same time will reduce the energy consumption and achieve green targets. In this respect, Operator-governed Opportunistic Networks (ONs), which are dynamically created, temporary, coordinated extensions of the infrastructure, are the basic constituents in the proposed approach. In addition, Cognitive Management Systems (CMSs), which comprise self-management and learning capabilities, can be exploited for ensuring fast and reliable establishment of ONs, achieving efficiently the desired goals. This paper presents the concept of ONs and their representative scenarios, as well as an evaluation of indicative test cases as a proof of concept of the aforementioned approach. Indicative simulation results are presented, which yield the conditions in which the adoption of such a solution can lead to lower costs and management decisions with a "greener" footprint.**

## I. INTRODUCTION

It is a common truth that the majority of existing and newly coming applications are more likely to be provided in their mobile/wireless manifestation. This has mainly emerged as a result of the expansion in using wireless internet access, and accordingly of the increase of data traffic volumes received/sent by internet enabled mobile devices. Such an increasingly demanding landscape in the area of wireless/mobile networks motivates the quest for technological solutions that will offer increased efficiency in resource utilization and application QoS provisioning and at the same time exhibit lower transmission powers and energy consumption.

Framed within this statement, in this paper we elaborate on a solution *(Figure 1)* that integrates *Wireless Network Infrastructure*, on the one side, and extensions of the infrastructure-based wireless network called *Opportunistic Networks (ONs)*, on the other side.

The wireless infrastructure can refer to a heterogeneous network that is composed of several Radio Access Networks (RANs), which include base stations (in the general sense), a packet-based core network for connecting these RANs and a common network management system [1][2].

On the other hand, ONs can be characterised as operator-governed, temporary, coordinated extensions of the wireless network infrastructure. In particular, ONs are governed by operators through the provision of policies, e.g. upon resource usage, as well as context/profile information and knowledge, which is exploited for their creation and maintenance. They are dynamically created in places and at the time they are needed to deliver application flows to mobile users. Furthermore, since ONs are extensions of the infrastructure, they will comprise various devices/terminals, potentially organized in an infrastructure-less mode, as well as elements of the infrastructure per se.

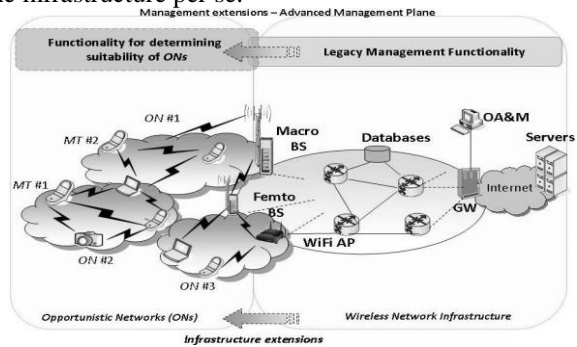

**Figure 1. High level architecture of our approach**

Furthermore, because of the highly dynamic nature of the environment, including traffic and applications issues, as well as the potential complexity of the infrastructure, a solution that incorporates self-management and learning mechanisms are deemed essential. Self-management enables a system to identify opportunities for improving its performance and adapting its operation without the need for human intervention. Learning mechanisms are important so as to increase the reliability of decision making. Learning mechanisms also enable proactive handling of problematic situations, i.e. identifying and handling issues that could undermine the performance of the system before these actually occur. In this respect, Cognitive Management Systems (CMSs), rendering both self-management and learning capabilities [3] seem appropriate for ensuring the fast and reliable establishment of ONs. CMSs can be located in both the network infrastructure and the terminals/devices. Moreover, it is envisaged that the coordination between CMSs and the exchange of information and knowledge can be provide by control channels. Such control channels may be logical channels transporting information on top of a physical network architecture.

The rest of the paper is structured as follows: Section II discusses the concept of CMSs for the management of ONs and their coordination with the infrastructure, while Section III presents scenarios of ONs along with their green footprint aspects. Section IV presents the simulation results with focus on the green achievements of the ONs. Also there is a presentation of the results obtained by simulations concerning green aspects. Finally, section V concludes the paper.

## II. COGNITIVE MANAGEMENT SYSTEM

CMSs comprise self-management and learning capabilities and can determine their behavior in an autonomous manner, reactively or proactively, according to goals, policies and knowledge. A CMS makes and enforces decisions on the creation of an ON by taking into account the context of operation (environment requirements and characteristics), goals and policies, profiles (related to applications, devices and users), and machine learning (for managing and exploiting knowledge and experience). The components of a CMS are depicted in Figure 2 and can be described as follows:

*Context*. The Context awareness functionality involves the monitoring of the status of the network, in order to be aware of the necessity to create an ON. In this sense, context includes traffic, radio and mobility conditions. In general, this part provides the characteristics of the environment.

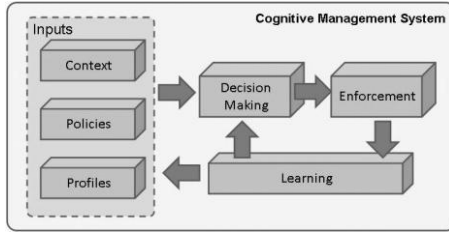

**Figure 2. Components of a CMS**

*Policies*. Policies designate rules that should be followed in context handling and refer to reconfiguration strategies that may applied by the opportunistic nodes, such as operator's preferences and priorities on goals to be achieved. These are related to the maximization of the QoS levels, and the minimization of cost factors (e.g. resource consumption).

*Profiles*. Profiles include preferences, requirements and constraints of user classes and applications which are required for the decision making. Particularly, regarding users the goal is to express their requirements, preferences, constraints. Regarding applications the goal is to express various options for their proper provision. Finally, regarding the served devices the goal is to describe their capabilities.

*Learning*. Learning taking place during the life-cycle of an ON, is used to produce knowledge on the acquired contextual and performance parameters, thus improving reliability and speed of management decisions.

The accuracy of the obtained knowledge on the context of the environment can be increased through the deployment of a cooperation mechanism between CMSs. Moreover, efficient coordination between the infrastructure and the devices in the scope of an ON is required, as well. Cognitive control channels enable such exchange of information and the coordination between CMSs. These control channels, depicted in Figure 3, can be based on the exploitation and evolution of two concepts: the Cognitive Pilot Channel (CPC) [4] and the Cognitive Control Radio (CCR) [5].

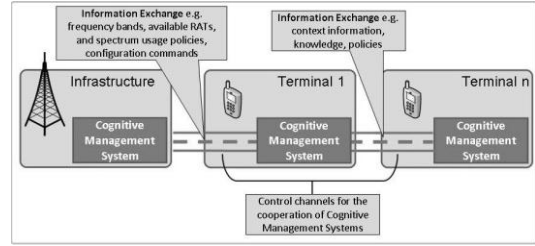

**Figure 3. Control Channel for the cooperation of CMS**

The CPC is defined as a channel (logical or physical) which conveys the elements of necessary information facilitating the operations of Cognitive Radio Systems and can be seen as an enabler for providing information from the network to the terminals, e.g., frequency bands, available RATs, and spectrum usage policies. The CCR can be seen as a channel for the peer-to-peer exchange of cognition related information between nodes (e.g., between terminals) belonging to the same network.

## III. OPPORTUNISTIC NETWORKS

### A. Scenarios

We present five representative scenarios, depicting different facets of Opportunistic Networks. In particular, the following scenarios are addressed: (i) opportunistic coverage extension, (ii) opportunistic capacity extension, (iii) infrastructure supported opportunistic ad hoc networking, (iv) opportunistic traffic aggregation in the radio access network and (v) opportunistic resource aggregation in the backhaul network.

In the coverage extension scenario, we consider a device like a laptop or a camera, which acts as a traffic source and is out of the coverage of the infrastructure. An ON is created in order to serve the out of coverage source with the use of an intermediate node. In the opportunistic capacity extension scenario, it is assumed that a specific area which experiences traffic congestion can be offloaded with the creation of an ON in order to re-route the traffic to non-congested Access Points (APs). The infrastructure supported opportunistic ad hoc networking scenario, considers closely located nodes with the same application interests, and the creation of an ON for traffic exchange.

The basic concept of traffic aggregation in the radio access network lies in the fact that there is a co-location of users in a certain service area that experience poor quality and therefore, an ON is created in order to enable traffic aggregation from the ON to the infrastructure through the gateway node. Finally the scenario of resource aggregation in the backhaul is represented by an ON that is created across multiple APs in order to primarily aggregate backhaul bandwidth and match the bandwidth of modern wireless access technologies towards the user with adequate bandwidth on the backhaul/ core network side.

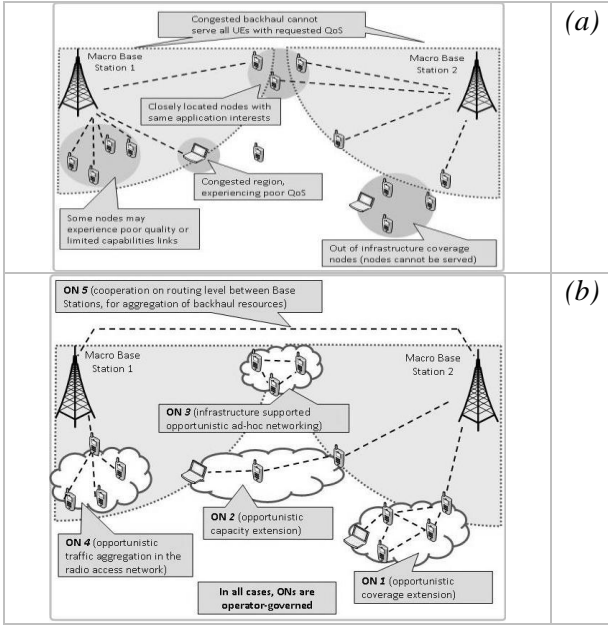

Figure 4. a) Problematic situations that may occur in a networking environment, b) Solutions by implementing the ON paradigm.

Figure 4 *(a)* depicts various situations which may occur in a network environment prior to the use of opportunistic networking, while Figure 4 *(b)* considers the solutions by implementing the ON paradigm.

### B. Green Footprint

In this subsection the Green Footprint of the five representative scenarios are presented.

In the coverage extension scenario of, lower transmission power level is achieved, which leads to lower electrical energy consumption. Therefore we achieve in this way higher green footprint and lower OPerational EXpenditures OPEX as well, which represents the expenses made by the operator for the operation of the infrastructure. Further analysis for the green aspects that arise from this scenario is presented in the following section.

In the coverage extension scenario the claim that the Opportunistic Networks influence the Green Footprint is strengthened, as they contribute to the reduction of the transmission power and consequently lead to lower electrical energy waste. Higher bit-rates are also possible, while capacity extension leads to less investment in infrastructure and consequently less hardware deployed. This means that lower OPEX, as well as CAPital EXpenditures CAPEX and the initial expenditure made, is achieved.

Moreover the Green Footprint of infrastructure supported opportunistic ad hoc networking is also proved by the energy consumption diminishment that is achieved through its application. In this case, higher bit rates are possible, as well. In particular, the application of closely located nodes leads to lower energy consumption due to the fact that there is no direct link in distance with the AP.

In the opportunistic traffic aggregation the green footprint is achieved by succeeding to support limited ON terminals, fact that leads to lower electrical energy consumption.

Finally opportunistic resource aggregation in the backhaul network leads to less investment in infrastructure and consequently less hardware deployed, which consequently results to higher green footprint and lower CAPEX.

In general the goal of the integration of ONs is twofold and it is summarized in the following:

- Provide light operations and migration to cost effective, secured, and deployable networking functionality, and
- Achieve the best ratio of performance to energy consumption and assure manageability.

## IV. EVALUATION

This section discusses on the simulations we have conducted for two of the aforementioned scenarios in order to investigate the green benefits and the overall network performance from the creation of the ON. We will also study performance metrics in order to assure that by achieving green performance we do not affect the overall network efficiency in the application provisioning.

### A. Scenario 1 – Coverage Extension

A set of scenarios and test cases were executed in the simulation environment, which was based on the widely used NS-2 simulator [9] and ran on an Intel Core i5 2.3 GHz with 4 GB of RAM and a 64-bit Operating System.

Without loss of generality, the topology comprises a single AP supporting IEEE 802.11g technology with a maximum offering data rate at 54Mbps. A set of 12 mobile terminals (MTs) are supported within the range of the AP, the four of which are selected to be the application consumers. VoIP application, based on the G.711 [10] voice encoding scheme for both the caller and the callee, is considered in the simulation scenarios exhibiting stringent resource requirements and real time sensitivity.

Describing the scenarios depicted in Figure 5, four phases (steps) are considered, each one corresponding to a specific percentage of the initial TRx power of the AP/MTs, namely: 100% (initial) / 100% (initial), 80% / 100%, 60% / 100% and 60% / 60% thus resulting in ranges R0/T0, R1/T0, R2/T0 and R2/T1, respectively. In general, during the reduction of the AP's TRx power, some of MTs are left out of the APs' range. These MTs are then supposed to create ONs with intermediate MTs in an ad-hoc manner and operate in WLAN 802.11g, as well. In particular, the initial transmission power of the AP and MTs is set equal to 0.03W and 0.02W, respectively.

Furthermore, each phase evolves also sub-phases depending on the mobility level of the intermediate MTs that are inside a predefined mobility domain using a random waypoint mobility model. We assumed 7 mobility levels (0m/sec – 15m/sec), with the same maximum speed for all the MTs that are inside this domain and pause times equal to 1 sec. Finally, for completion reasons, we also experiment with the routing protocols [11] namely, DSR, AODV, OLSR and GRP, which will be used to route traffic to MTs that are found out of range during the AP's and MTs' transmission power reduction.

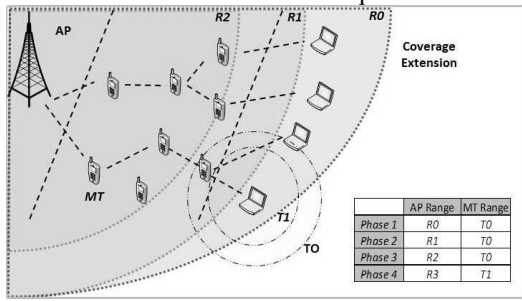

**Figure 5. Simulation scenarios**

Figure 6 depicts the total power consumption that is required for supporting the data traffic in the scenario prior *(phase 1)* and after the formation of the ON *(phase 2-4)*. All the nodes in the network and the infrastructure element *(AP)* are considered in the computation of the total energy consumption. As depicted, the formation of the ON can result in a reduction of 40% of the initial transmission power. This result supports our initial claims of better use of resources in terms of energy consumption.

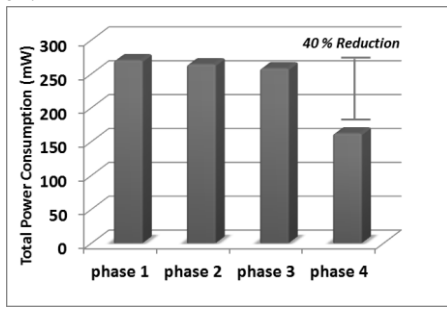

**Figure 6. Total Power Consumption in mW.**

In the sequel, we also focus on specific QoS metric, which is used to evaluate conditions and assist in coming up with useful recommendations with respect to the creation of the ONs networks with the possible green gains. We focus herewith on performance metric associated with the QoS levels that the applications will be provided through the ON(s), namely Application Delay (sec).

Figure 7 depicts the end-to-end delay that VoIP suffers averaged on the application MTs. The single dotted line corresponds to the maximum acceptable delay. Generally, as the mobility level increases, the overall delay also increases. This is due to the fact that the mobility in the intermediate nodes, can significantly impact the performance of the ad-hoc routing protocols, including the packet delivery ratio, the control overhead and the data packet delay [12].

Moreover as the number of intermediate nodes increases, while the AP's range is shrinking, the overall delay also increases since more MTs participate in routing and forwarding of the received packets. In the same figure, the existed solutions in terms of mobility level are depicted with the anticipated green benefits. In *phase 2 and 3*, it is observed that all mobility levels has acceptable values and can result to a *20% and 40%* reduction of transmission power, respectively.

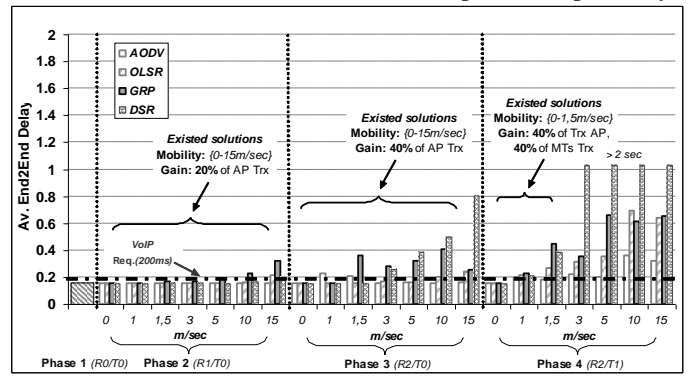

**Figure 7. Average End to End Delay (sec) per Node in 4 Phases**

On the other hand in phase 4 there are only three acceptable solutions (*0-1,5*m/sec) and can result in a reduction of 40% of the required transmission power. Therefore, there is some degree of freedom regarding the creation of the ON to be applied, which can vary depending on the level of emphasis given to specific target QoS levels and aimed savings in transmission power.

In general, as simulations showed, AP/MTs transmission power reductions of 60% can be done without impacting too much the service provisioning. This means that the supported application remains at acceptable QoS levels with less AP/MT power resources. Therefore this reduction will result important, savings in the total transmission power in the network.

### B. Scenario 2 – Opportunistic traffic aggregation in the radio access network

Focusing on the traffic aggregation scenario, there may be users that face poor channel quality towards the infrastructure, because they may residing at the edge of the AP and at the same time very good channel quality towards some of their neighbors. It is obvious that users with poor channel quality, compared to those with better quality, need more resources (e.g. power, time) to transmit the same amount of data. After the creation of the ON, users with good channel conditions towards the infrastructure will be responsible for forwarding traffic to those that have poor channel conditions, through their direct interfaces. Therefore, the ON will increase the overall system capacity and resource utilization, and offer a service in an energy efficient manner.

For the simulations, a set of four nodes set-up direct connections (via cellular interfaces) with a network in order to transmit data packets of the size of *1*MB each. It is assumed that users' devices are equipped with 3G interfaces (for the connection with the AP) and with IEEE 802.11g interfaces for the peer-to-peer connections among them. At some point in time the quality of the nodes' connections significantly drops, thus the connection throughput is limited for three of them to

*0.5*Mbps. At the same time, these nodes maintain very good channel quality towards some of their neighbors, offering a rate of *54*Mbps by using IEEE 802.11g interface. The AP in collaboration with the nodes, detects that situation, and initiates the process for the creation of an ON. They jointly determine one or several nodes, with high channel conditions to aggregate and/or relay traffic of other users, which have poor channel conditions towards the infrastructure.

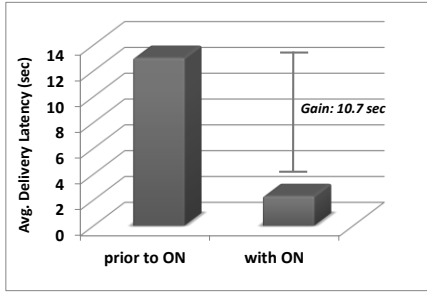

**Figure 8. Average Delivery Latency Estimation in sec**

Figure 8, depicts the anticipated average delivery latency prior to and after the formation of ON, as measured in the ONE simulator. It corresponds to the one-way time (in seconds) from the source sending a packet to the destination receiving it. With direct links, it is observed that the average delivery latency is around *13*sec, while the deployment of the ON yields a significant drop of this metric to *2.3*sec.

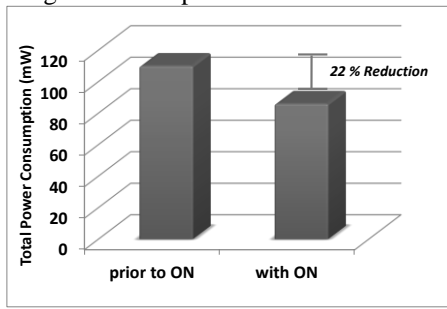

**Figure 9. Total Power Consumption in mW.**

Moreover, Figure 9 depicts the total power consumption that is required for the traffic of the scenario prior and after the creation of the ON. It considers again all nodes in the range of the AP and the infrastructure element. As Figure 9 depicts, the creation of the ON can result in a reduction of 22% of the required transmission power, which can be justified by the shorter direct links that are used for forwarding traffic within the ON. This result in this scenario also proves our claims of better use of resources in terms of energy consumption. Finally, the CMSs can ensure fast and reliable establishment of ONs and perform well when facing same situations, thus resulting in faster reductions of the energy consumption, achieving efficiently green targets.

## V. CONCLUSIONS

The enormous increase in wireless access demand requires efficient solutions for problems, which are associated with increased efficiency in resource utilization and application QoS provisioning, as well as lower energy/power consumption and reduced total cost. To this effect, this paper covers a proposed solution that encompasses ONs and CMSs for managing the ONs. It is claimed that the proposed ON-based solution can prove beneficiary achieving green targets in various scenarios, and this is also strengthened through a set of indicative simulations and results.

## ACKNOWLEDGMENT

This work is performed in the framework of the European-Union funded project OneFIT (www.ict-onefit.eu). The project is supported by the European Community's Seventh Framework Program (FP7). The views expressed in this document do not necessarily represent the views of the complete consortium. The Community is not liable for any use that may be made of the information contained herein. This work is also supported by the Cost Action IC0902, "Cognitive Radio and Networking for Cooperative Coexistence of Heterogeneous Wireless Networks". Finally this work supports training activities in the context of the ACROPOLIS (Advanced coexistence technologies for Radio Optimisation in Licenced and Unlicensed Spectrum -Network of Excellence) project (http://www.ict-acropolis.eu).

## REFERENCES

[1] IEEE Std 1900.4™-2009, IEEE Standard for Architectural Building Blocks Enabling Network-Device Distributed Decision Making for Optimized Radio Resource Usage in Heterogeneous Wireless Access Networks," Jan. 2009

[2] ETSI, TR 102.838, V1.1.1 (Oct. 2009), "Reconfigurable Radio Systems (RRS); Summary of feasibility studies and potential standardization topics", 2010

[3] R. Thomas, D. Friend, L. DaSilva, A. McKenzie, "Cognitive networks: adaptation and learning to achieve end-to-end performance objectives", IEEE Commun. Mag., Vol. 44, No. 12, pp. 51-57, Dec. 2006

[4] ETSI TR 102.683, v1.1.1, "Reconfigurable Radio Systems (RRS); Cognitive Pilot Channel (CPC)", 2009

[5] ETSI TR 102.802, v1.1.1., "Reconfigurable Radio Systems (RRS); Cognitive Radio System Concepts", 2010

[6] L. Pelusi, A. Passarella, M. Conti. "Opportunistic networking: Data forwarding in disconnected mobile ad hoc networks." IEEE Communications Magazine, vol. 44, pp. 134–141, Nov 2006

[7] L. Lilien, A. Gupta, Z. Yang, "Opportunistic networks for emergency applications and their standard implementation framework", in Proc. of IEEE International Performance, Computing, and Communications Conference,(IPCCC 2007), 2007.

[8] C.-M. Huang, K.-C. Lan, C.-Z. Tsai, "A survey of opportunistic networks", in Proc. of the 22nd International Conference on Advanced Information Networking and Applications - Workshops, (AINAW 2008), Okinawa, Japan, March 2008

[9] Network Simulator 2 (NS-2), http://www.isi.edu/nsnam/ns/

[10] ITU-T recommendation G.711, Aspects of digital transmission Systems

[11] Charles E. Perkins. Ad Hoc Networking. Addision Wesley, 2001

[12] T. Camp, J. Boleng, and V. Davies. A Survey of Mobility Models for Ad Hoc Network Research. Wireless Communications & Mobile Computing (WCMC): Special issue on Mobile Ad Hoc Networking: Research, Trends and Applications, 2(5):483–502, 2002